# *ASCA* observations of the Seyfert 1 galaxies Mrk 1040 and MS 0225.5+3121

C. S. Reynolds[1], A. C. Fabian[1] and H. Inoue[2]
[1] *Institute of Astronomy, Madingley Road, Cambridge CB3 OHA*
[2] *Institute of Space and Astronautical Science, Yoshinodai, Sagamihara, Kanagawa 229, Japan*

27 April 1995

**ABSTRACT**
We present *ASCA* observations of the Seyfert 1 galaxies Mrk 1040 and MS 0225.5+3121. Mrk 1040 was found to have decreased in flux by almost a factor of 4 since an *EXOSAT* observation 10 years ago. The energy spectrum of Mrk 1040 displays complexity both at soft energies (below 0.8 keV) and at hard energies (6–7 keV). The latter is readily interpreted as fluorescent K$\alpha$ emission from cold iron expected when the primary X-ray source illuminates cold optically-thick material. This line is both broad (with FWHM 16 000–70 000 km s$^{-1}$) and strong (equivalent width $\sim 550 \pm 250$ eV) suggesting that it originates from material close to the compact object with non-solar abundances. Abundance effects on the equivalent width of such a line are investigated via Monte Carlo simulations. We find that strong lines can be produced with physically plausible abundances. The effect of abundances on the associated reflection continuum is also discussed. The soft spectral complexity implies either a strong soft excess together with intrinsic absorption, or a complex absorber. Various models for the nature of such a complex absorber are discussed. MS 0225.5+3121 shows no evidence for any variability and has a spectrum that is well described by a power law with Galactic absorption.

## 1 INTRODUCTION

X-ray observations of active galactic nuclei (AGN) probe the innermost regions of the central engine. Temporal and spectral studies in the X-ray band can provide direct information on the physical state and geometry of matter in AGN as well as indicate the primary high-energy emission mechanisms. Much of our understanding in this field has come from the investigation of nearby, X-ray bright, Seyfert 1 galaxies.

To a first approximation, the energy spectrum of these objects is described by a power law. Deviations from a power law are interpreted as being due to X-ray reprocessing by matter in the immediate environment of the central engine. *EXOSAT* and *Ginga* observations of nearby Seyfert 1 galaxies found evidence for the fluorescent K-shell emission (at 6.4 keV) and absorption from cold iron (Nandra et al. 1989; Matsuoka et al. 1990) expected from the X-ray illumination of cold, dense material in the central engine (Guilbert & Rees 1988; Lightman & White 1988; George & Fabian 1991; Matt, Perola & Piro 1991). In addition, *Ginga* found spectral flattening above $\sim 10$ keV associated with the accompanying reflected continuum (Nandra, Pounds & Stewart 1990; Nandra & Pounds 1994). *Ginga* also found many Seyfert 1 galaxies to have an absorption edge corresponding to partially-ionized iron (Nandra & Pounds 1994). This was taken to be evidence for partially-ionized, optically-thin material along the line of sight to the central X-ray source, material that has become known as the warm absorber (Halpern 1984). Theoretical aspects of such material are discussed by Netzer (1993), Krolik & Kriss (1995) and Reynolds & Fabian (1995). Clear evidence for the warm absorber was provided by *ROSAT* Position Sensitive Proportional Counter (PSPC) spectra which show the presence of warm oxygen absorption edges (Nandra & Pounds 1992).

*ASCA* (Tanaka et al. 1994) has made major advances in the study of AGN (Matsuoka 1994). The superior spectral energy resolution of the solid-state imaging spectrometer (SIS) allows X-ray spectral features to be accurately identified and measured. *ASCA* observations of MCG$-$6$-$30$-$15 clearly resolved and separated the O VII and O VIII K-shell absorption edges, thereby confirming the warm absorber hypothesis (Fabian et al. 1994). The warm absorber was also seen to vary both between observations and during individual observations (Fabian et al. 1994; Reynolds et al. 1995). This short term variability, together with the lack of *flux-correlated* changes in the inferred ionization state of the warm absorber, requires us to abandon simple one-zone equilibrium models for the absorbing matter. Instead, we must consider inhomogeneous and/or dynamic, non-equilibrium warm absorber models (Reynolds & Fabian 1995; Reynolds et al. 1995; see also Krolik & Kriss 1995).

Mrk 1040 (NGC 931; z=0.016) is a Seyfert 1 galaxy included in the *EXOSAT* spectral survey of Turner & Pounds (1989). The *EXOSAT* spectrum was well described by a power law with photon index $\Gamma = 1.72 \pm 0.21$ and a cold absorbing column density of $3.5^{+2.7}_{-1.9} \times 10^{21}$ cm$^{-2}$ (the Galactic absorption in the direction of Mrk 1040 is $7.2 \times 10^{20}$ cm$^{-2}$; Elvis, Lockman & Wilkes 1989). Significant absorption from the interstellar medium (ISM) of the host galaxy may be expected due to the high inclination of the host galaxy (Amram et al. 1992). The inferred intrinsic (i.e. unabsorbed) 2–10 keV luminosity is $2.3 \times 10^{43}$ erg s$^{-1}$ for the *EXOSAT* observation (taken in 1984; throughout this paper it is assumed that $H_0 = 50$ km s$^{-1}$ Mpc$^{-1}$ and $q_0 = 0$). Mrk 1040 was also observed by the *Einstein* Imaging Proportional Counter (IPC) in 1980 and displayed a similar luminosity (Wilkes et al. 1994). No pointed *Ginga* or *ROSAT* observations of this source have been made. *ASCA* observations were proposed in order to investigate details of the low-energy absorption.

MS 0225.5+3121 was discovered by the *Einstein* Observatory Extended Medium Sensitivity Survey and optically identified as a Seyfert 1 galaxy (z=0.058) by Stephens (1989). Being ∼15 arcmins from Mrk 1040, it is in the same *ASCA* gas imaging spectrometer (GIS) field of view.

The present paper describes an analysis of *ASCA* observations of Mrk 1040 and MS 0225.5+3121. Section 2 briefly describes the observations. Section 3 presents the temporal and spectral analysis for Mrk 1040. We demonstrate the presence of spectral complexities both at soft energies ($< 1$ keV) and hard energies (between 6–7 keV). Possible interpretations of these complexities are discussed. The investigation of MS 0225.5+3121 is given in Section 4. Section 5 provides a summary of the important results and implications.

## 2 OBSERVATIONS

Mrk 1040 and MS 0225.5+3121 were observed by *ASCA* on 1994 19/20 August. Both SIS and GIS data were obtained for Mrk 1040 whereas only GIS data were obtained for MS 0225.5+3121 due to the restricted SIS field of view. The SIS data were cleaned in order to remove the effects of hot and flickering pixels and subjected to the following data-selection criteria : i) the satellite should not be in the South Atlantic Anomaly (SAA), ii) the object should be at least 7 degrees above the Earth's limb, iii) the object should be at least 25 degrees above the day-time Earth limb and iv) the local geomagnetic cut-off rigidity (COR) should be greater than 6 keV. The GIS data were cleaned to remove the particle background and subjected to the following data-selection criteria : i) the satellite should not be in the SAA, ii) the object should be at least 7 degrees above the Earth's limb and iii) the COR should be greater than 7 keV. SIS and GIS data that satisfy these criteria shall be referred to as 'good' data.

After the above data selection, there are 18 900 s of good data per SIS detector and 20 700 s of good data per GIS detector. Images, light curves and spectra were then extracted from this good data for each of the four instruments (two SIS and two GIS). To enable $\chi^2$ fitting, spectra are binned so as to have a minimum of 20 photons per energy bin (thereby ensuring Gaussian photon statistics in each energy bin).

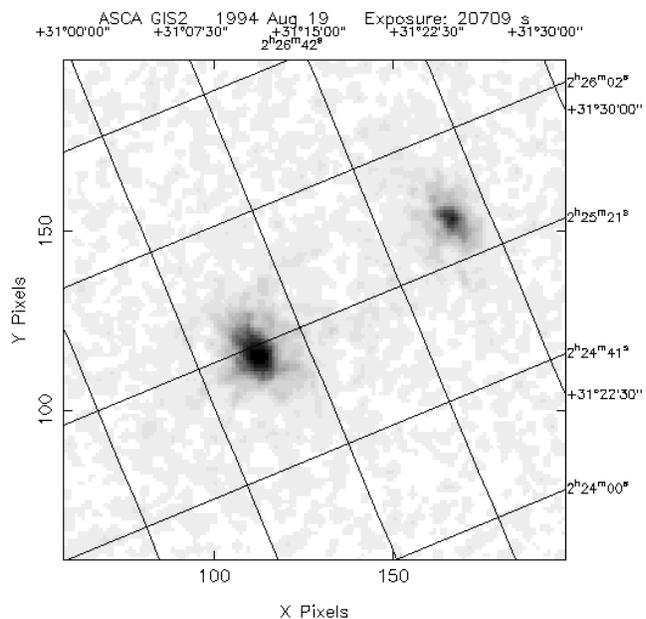

**Figure 1.** Integrated GIS2 image (20 700 s of data). Emission from Mrk 1040 (lower-left) and MS 0225.5+3121 (upper-right) is clearly detected and is consistent with point-like sources. The co-ordinates are equinox 2000.

Fig. 1 shows the GIS2 image for this observation. Mrk 1040 and MS 0225.5+3121 are clearly detected. Furthermore, these are the only sources in the GIS field of view with a positive detection. There is no evidence for any extended emission from either of these sources. Table 1 gives *average* values for the SIS count rate, GIS count rate, background count rates, 2–10 keV flux and 2–10 keV luminosity for both Mrk 1040 and MS 0225.5+3121. For both the SIS and GIS, counts have been extracted from a region of 5 arcmins radius centred on the appropriate source. Only a negligible fraction of the source counts will lie outside of this extraction radius.

## 3 TEMPORAL AND SPECTRAL ANALYSIS OF MRK 1040

### 3.1 Temporal analysis

No evidence was found for variability in the flux from Mrk 1040 during the 12 hr period of the *ASCA* observation. However, a comparison of Table 1 with the results of Turner & Pounds (1989) reveals a significant decline in the 2–10 keV luminosity from $2.3 \times 10^{43}$ erg s$^{-1}$ (epoch 1984) to $5.9 \times 10^{42}$ erg s$^{-1}$ (epoch 1994) [Note that both of these average luminosities are determined to within 2 per cent.]

### 3.2 Spectral analysis

Spectra for Mrk 1040 were extracted from each of the four instruments on board *ASCA*. Background spectra were taken from source free regions of the same field of view. For each SIS, the background data were taken from a 45 arcmin$^2$ region towards the edges of the chip. For each GIS, the background data were taken from a 450 arcmin$^2$ source free region. All counts were corrected for vignetting. Spectral fitting of various models was then performed using all four background subtracted spectra simultaneously.

| | Observational Parameter | Mrk 1040 | MS 0225.5+3121 |
|---|---|---|---|
| | SIS0 source count rate (cts s$^{-1}$) | $0.133 \pm 0.003$ | no SIS data |
| | GIS2 source count rate (cts s$^{-1}$) | $0.111 \pm 0.002$ | $0.039 \pm 0.001$ |
| | SIS0 background count rate ($10^{-4}$ cts arcmin$^{-2}$ s$^{-1}$) | $6.5 \pm 0.3$ | $6.5 \pm 0.3$ |
| | GIS2 background count rate ($10^{-4}$ cts arcmin$^{-2}$ s$^{-1}$) | $1.74 \pm 0.04$ | $1.7 \pm 0.04$ |
| | $F_{2-10\,\mathrm{keV}}$ ($10^{-12}$ erg cm$^{-2}$ s$^{-1}$) | $5.1 \pm 0.1$ | $1.8 \pm 0.1$ |
| | $L_{2-10\,\mathrm{keV}}$ ($10^{42}$ erg s$^{-1}$) | $5.9 \pm 0.1$ | $25 \pm 2$ |

**Table 1.** Average SIS0 count rates, GIS2 count rates, the 2–10 keV flux $F_{2-10\,\mathrm{keV}}$ and the 2–10 keV luminosity $L_{2-10\,\mathrm{keV}}$ (*not* corrected for absorption) for the *ASCA* observation of Mrk 1040 and MS 0225.5+3121. For Mrk 1040, spectral model B (from Table 2) is used to compute the flux and luminosity. For MS 0225.5+3121, the best-fitting spectral model of Section 4.2 is used.

| | Model | Best-fitting model parameters | $\chi^2$/dof |
|---|---|---|---|
| A | PL + Gal ABS | $\Gamma = 1.20 \pm 0.03$ | 529/371 |
| B | PL + Gal ABS +ABS | $N_\mathrm{H} = 2.6^{+0.7}_{-0.4} \times 10^{21}$ cm$^{-2}$<br>$\Gamma = 1.54^{+0.07}_{-0.04}$ | 384/370 |
| C | PL + Gal ABS + ABS + BB | $N_\mathrm{H} = 3.6 \pm 0.8 \times 10^{21}$ cm$^{-2}$<br>$\Gamma = 1.63^{+0.07}_{-0.08}$<br>$k_\mathrm{B}T = 66^{+11}_{-36}$ eV | 369/368 |
| D | PL + Gal ABS + par ABS | $\Gamma = 1.66 \pm 0.08$<br>$N_\mathrm{H} = 5.3 \pm 1.3 \times 10^{21}$ cm$^{-2}$<br>$f = 0.81 \pm 0.06$ | 367/366 |
| E | PL + Gal ABS + var ABS | $\Gamma = 1.60^{+0.08}_{-0.07}$<br>$N_\mathrm{H} = 1.2^{+0.4}_{-0.3} \times 10^{21}$ cm$^{-2}$<br>$Z = 15 \pm 4$ | 357/366 |
| F | PL + Gal ABS + WARM | $\Gamma = 1.54^{+0.06}_{-0.04}$<br>$N_\mathrm{W} = 4.4^{+0.7}_{-0.9} \times 10^{21}$ cm$^{-2}$<br>$\xi = 9.0^{+5.0}_{-3.5}$ erg cm s$^{-1}$ | 355/366 |

**Table 2.** Results of spectral fitting for *ASCA* data of Mrk 1040. Fitting was performed to data from all four *ASCA* detectors simultaneously. Abbreviations are as follows; PL=power law (photon index $\Gamma$), BB=Black-body (temperature T), Gal ABS=Galactic (neutral) absorption, ABS=intrinsic neutral absorption with solar abundances (column density $N_\mathrm{H}$), par ABS=intrinsic neutral absorption with only partial covering of the primary source (covering fraction $f$), var ABS=intrinsic neutral absorption with the abundance of iron/nickel as a free parameter (hydrogen column density $N_\mathrm{H}$, iron/nickel abundance $Z$ relative to solar) and WARM=intrinsic ionized ('warm') absorber (column density $N_\mathrm{W}$ and ionization parameter $\xi$). All errors are quoted at the 90 per cent confidence level for 1 interesting parameter, $\Delta\chi^2 = 2.7$.

Table 2 summarizes the results of the spectral fitting for Mrk 1040. A model consisting of a simple power law modified by Galactic cold absorption (assumed to be of solar abundances) is a poor description of the data with $\chi^2 = 529$ for 371 degrees of freedom (dof). Allowing the column density of cold absorbing material to be a free parameter leads to a highly significant improvement (at more than 99.9 per cent confidence) in the goodness of fit ($\chi^2 = 384$ for 370 dof) and indicates a large amount of intrinsic absorption within Mrk 1040. The derived photon index and cold column density are consistent with the *EXOSAT* result of Turner & Pounds (1989). Fig. 2 shows the data compared with this model (model B in Table 2). It can be seen that additional (unmodelled) spectral complexity exists both at soft energies (below 0.8 keV) and at hard energies (6–7 keV).

### 3.2.1 *Soft spectral complexity*

The failure of the simple absorbed power law (model B) to explain the soft spectrum leads us to consider more complicated models. Detecting absorption edges would be a valuable way of investigating the nature of an intrinsic absorber. Unfortunately, the unexpectedly low-luminosity of Mrk 1040 (and thus the reduction in the number of photons detected) prevents us from clearly defining such absorption edges. Instead, we must attempt to model the general behaviour of the soft spectrum.

We consider four additional models (detailed in Table 2). The first (model C) consists of a power law spectrum with a soft excess (modelled as a black-body) absorbed by neutral material. The other three models (D, E and F) all assume only a power law primary spectrum modified by var-ious types of absorption. Model D supposes absorption by neutral material that only covers a fraction $f$ of the primary X-ray source, model E has absorption by neutral material with variable iron abundance and model F has a warm (ionized) absorber (using the warm absorber models of Fabian et al. 1994 and Reynolds & Fabian 1995). Table 2 demonstrates that model F best describes the data ($\chi^2 = 355$ with 366 dof). Models C, D and E also provide good fits to the data with $\chi^2$=369, 367 and 357 respectively (for 366 dof each). Each of four these models is a statistically significant improvement at the 99.9 per cent level (according to the F-test; Bevington 1969). However, there is no statistical preference between models.

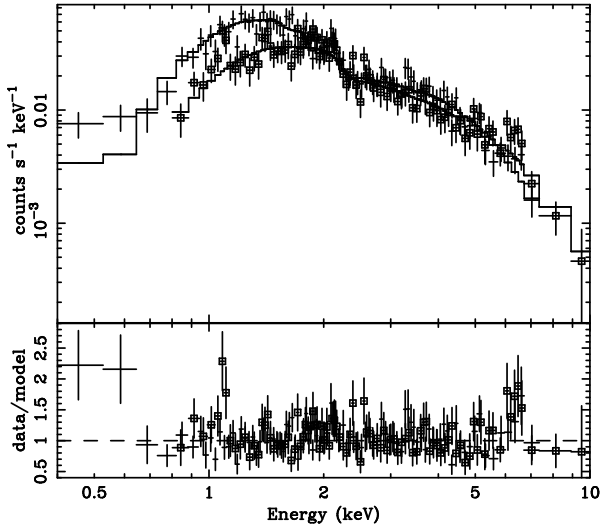

**Figure 2.** *ASCA* SIS0 (plain crosses) and GIS2 (crosses with squares) spectrum for Mrk 1040. SIS1 and GIS3 were used for spectral fitting but have been omitted for clarity of presentation. $1-\sigma$ errors are shown. Also shown is a model consisting of a primary power law with photon index $\Gamma = 1.54$ absorbed by neutral material with column density $N_H = 3.3 \times 10^{21}$ cm$^{-2}$. Subtracting the absorbing column expected from the Galaxy, an intrinsic absorbing column density of $N_H = 2.6 \times 10^{21}$ cm$^{-2}$ is inferred. The bottom panel shows the deviations of the data from the model. Note the discrepancies between the model and the data below 0.8 keV and between 6–7 keV. These are extensively discussed in the text.

### 3.2.2 *Hard spectral complexity : the iron line*

Spectral complexity in the 6–7 keV range is expected from the fluorescent K-shell emission of iron. To model such an emission line, a Gaussian with mean energy $E$ and width $\sigma$ (in the rest frame of Mrk 1040) was added to the models B, C, D, E and F of Table 2. The resulting spectral fits are shown in Table 3. We note that the energy resolution of the *ASCA* SIS at these energies is $\sim 150$ eV.

These fits show that the addition of such a line is highly significant according to the F-test (Bevington 1969), with $\Delta\chi^2 > 20$ in each case (for three additional degrees of free-dom). Such a reduction in $\chi^2$ is significant at the 99 per cent level (Bevington 1969). Irrespective of the base model used, the line energy is in good agreement with that expected from the K-shell fluorescence of cold iron. The addition of the line also increases the best-fitting photon index $\Gamma$ to values more typical of other Seyfert galaxies. Another robust result is that the line is broad, with $\sigma$ constrained to lie within the range $\sim 0.14$–0.60 keV. If due to Doppler shifts, the corresponding FWHM velocities would be 16 000–70 000 km s$^{-1}$. This is more than an order of magnitude greater than optical/UV broad lines which have a FWHM of $\sim 1\,600$ km s$^{-1}$ (Osterbrock & Shuder 1982). These large velocities imply that the X-ray line emitting material lies close to the central compact object, probably in the form of an accretion disk. The line is strong with an equivalent width $\sim 550 \pm 250$ eV (a slab of material with solar abundances, as defined by Morrison & McCammon 1983, subtending $2\pi$ steradians at the X-ray source would give rise to a fluorescence iron line with EW$\sim$150 eV; George & Fabian 1991). Taken at face value, this would imply iron abundances significantly in excess of the solar value, a possibility which is discussed below in detail.

The possible presence of a reflection continuum has also been investigated. This would be expected to create a spectral flattening at high end of the *ASCA* band. However, the addition of such a reflected continuum was not a significant improvement ($\Delta\chi^2 < 3$) and did not affect the other parameters of the spectral fitting in any significant way.

### 3.3 Discussion

#### 3.3.1 *Cold and warm absorption*

The soft complexities in the *ASCA* spectrum of Mrk 1040 lead us to consider one of the four following scenarios;

(i) The presence of a strong soft excess with a intrinsic neutral absorbing column density of $N_H = 3.6^{+0.7}_{-0.4} \times 10^{21}$ cm$^{-2}$ (model C).

(ii) A high column density ($N_H = 5.3 \pm 1.3 \times 10^{21}$ cm$^{-2}$) of absorbing neutral material covering only $f = 0.81 \pm 0.06$ of the primary X-ray source (model D).

(iii) Absorption by neutral material ($N_H = 1.2^{+0.4}_{-0.3} \times 10^{21}$ cm$^{-2}$) which is over-abundant in iron/nickel by a factor of $Z = 15 \pm 4$ relative to the lighter elements (model E).

(iv) Absorption by ionized material (i.e. warm absorption) with column density $N_W = 4.4^{+0.7}_{-0.9} \times 10^{21}$ cm$^{-2}$ and ionization parameter $\xi = 9.0^{+5.0}_{-3.5}$ erg cm s$^{-1}$ (model F).

Each of these models provides a (statistically) satisfactory description of the present *ASCA* data. Thus, these data alone are not able to distinguish between these models. To proceed further we must examine the constraints imposed by observations at other wavelengths and investigate the physical plausibility of the various models.

Significant quantities of neutral gas along the line of sight to the active nucleus may reveal its presence through the reddening effect of associated dust at UV/optical wavelengths. The broad-line Balmer decrement H$\alpha$/H$\beta$ for Mrk 1040 is 6.2 (Rudy 1984). Assuming case-B recombination, the intrinsic (unreddened) Balmer decrement is 2.85 and corresponding extinction (Ward et al. 1987) is E(B$-$V)=0.63. The further assumption that the gas-to-dust ratio takes on the local (Galactic) value gives a column den-sity of $N_{\rm H} \sim 3.5 \times 10^{21}$ cm$^{-2}$ (Burstein & Heiles 1978). Thus, the large Balmer decrement is consistent with significant intrinsic absorption by cold gas at the level of 3–4 $\times 10^{21}$ cm$^{-2}$. This would tend to support models C and D (we note in passing that the strong big blue bump implied by model C does not violate the observed UV limits). However, it is recognized that the broad-line Balmer decrement is a poor indicator of the reddening due to its uncertain intrinsic (unreddened) value. The effects of collisional excitation and strong X-ray heating in the broad-line emitting gas can change the intrinsic Balmer decrement from the case-B value towards the value appropriate for pure collisional excitation (Gaskell & Ferland 1984). More seriously, there are now a number of objects known in which the intrinsic absorption expected on the basis of the observed Balmer decrement is inconsistent with constraints imposed by the X-ray data (e.g. Ark 564, Brandt et al. 1994; MCG−6−30−15, Pineda et al. 1980). Possible resolutions of this discrepancy are discussed in Brandt et al. (1994).

| Base Model | $\Gamma$ | $E$ (keV) | $\sigma$ (keV) | EW (eV) | $\chi^2$/dof |
|---|---|---|---|---|---|
| B | $1.64 \pm 0.08$ | $6.41 \pm 0.15$ | $0.29^{+0.18}_{-0.15}$ | $500 \pm 200$ | 361/367 |
| C | $1.75 \pm 0.11$ | $6.41 \pm 0.15$ | $0.33^{+0.20}_{-0.15}$ | $550 \pm 250$ | 342/365 |
| D | $1.80 \pm 0.10$ | $6.41 \pm 0.14$ | $0.35^{+0.20}_{-0.15}$ | $650^{+250}_{-300}$ | 338/363 |
| E | $1.71 \pm 0.10$ | $6.41 \pm 0.17$ | $0.33^{+0.22}_{-0.16}$ | $580^{+200}_{-220}$ | 332/363 |
| F | $1.66^{+0.11}_{-0.10}$ | $6.42 \pm 0.17$ | $0.35^{+0.25}_{-0.17}$ | $600^{+300}_{-250}$ | 341/363 |

**Table 3.** Results of adding a Gaussian emission line to the spectral fits of Table 2. Shown here is the base model (i.e. model from Table 2 to which Gaussian line was added), $\Gamma$ for the underlying power law of the new best-fitting model, line energy $E$, line width $\sigma$, equivalent width EW and the goodness of fit. All errors are quoted at the 90 per cent confidence level for one interesting parameter, $\Delta\chi^2 = 2.7$.

Absorption by cold material with greater than solar iron abundance (model E) can reproduce the observed spectrum due to the strong resultant iron L-edges below 1 keV. This model requires iron to be enhanced by a factor of $15 \pm 4$ relative to the light metals. Such an enhancement is consistent with the strong iron K$\alpha$ emission line of section 3.2.2 and is discussed in detail in the next section.

Constraints on the warm absorber model (model F) of Mrk 1040 can be derived by considering the UV line fluxes expected. The photoionization code CLOUDY (Ferland 1992) has been used to examine the physical state of photoionized material with the parameters given in Table 2. We find that the material would have a temperature of $\sim 40\,000$ K with C IV $\lambda 1549$ and O VI $\lambda 1035$ emission being the dominant cooling processes. Assuming the warm material to have unit covering fraction, the corresponding line fluxes would be $\sim 1 \times 10^{-12}$ erg cm$^{-2}$ s$^{-1}$ and $\sim 4 \times 10^{-12}$ erg cm$^{-2}$ s$^{-1}$ respectively. Examination of the *IUE* spectrum (Courvoisier & Paltani 1992) of Mrk 1040 fails to show any such C IV $\lambda 1549$ line (with an upper limit of $6 \times 10^{-15}$ erg cm$^{-2}$ s$^{-1}$ Å$^{-1}$). However, there are several ways in which a null *IUE* detection of this line can be made compatible with the presence of a warm absorber. First, if the warm absorbing material resided within the broad line region (BLR), Doppler broadening of these emission lines could prevent their detection against the continuum source. Material velocities of $\sim 5\,000$ km s$^{-1}$ or more would be required to render the C IV line undetectable in the *IUE* observation. This is significantly greater than the FWHM of the observed optical/UV broad lines ($\sim 1\,600$ km s$^{-1}$; Osterbrock & Shuder 1982). Secondly, the warm material may contain dust (provided it is sufficiently far from the continuum source to avoid photo-destruction or melting). Ultra-violet extinction might then suppress the observed line emission as well as explain the high (broad-line) Balmer decrement. Thirdly, the covering fraction of the warm material may be small thereby giving the observed absorption features without much associated emission. Lastly, the warm material may not be photoionization dominated and/or be far from equilibrium (Reynolds & Fabian 1995; Reynolds et al. 1995). This would drastically alter the intrinsic emission line spectrum (Krolik & Kriss 1995).

### 3.3.2 *Emission and absorption from iron*

Section 3.2.2 examines the apparent spectral complexity between 6–7 keV and shows that it is well described by a broad Gaussian emission line centered at 6.4 keV (in the rest frame of Mrk 1040). This is interpreted as being the K$\alpha$ fluorescence line of iron in a low ionization state resulting from the reflection of primary X-rays from cold material. The FWHM of the line is 16 000-70 000 km s$^{-1}$ thereby leading us to conclude that the line photons originate near to the central compact object (e.g. in an accretion disk). A slab of cold material with solar abundances (as defined by Morrison & McCammon 1983) subtending $2\pi$ at the X-ray source would give rise to a fluorescence line with equivalent width $\sim 150$ eV (George & Fabian 1991). However, the observed line has a much greater equivalent width ($550 \pm 250$ eV). Here we investigate the affect of non-solar abundances on the equivalent width of this line and the associated reflection continuum over a wider abundance range than that considered by George & Fabian (1991) and Matt, Perola & Piro (1991).

We address this issue using a Monte Carlo simulation of X-ray reflection similar to that used by George & Fabian (1991). Photons are randomly assigned an initial energy (according to a power law spectrum with $\Gamma = 1.9$) and an incident direction (corresponding to an isotropic source). Photon paths are followed until they escape the slab (i.e. are reflected) or are absorbed. The model includes Compton scattering and the K-shell photoelectric absorption and fluorescence of C, O, Ne, Mg, Si, S, Ar, Ca, Cr, Fe and Ni. Only K$\alpha$ are considered with the exception of iron for which K$\alpha$ and K$\beta$ are included. Photoelectric cross sections are taken from Verner et al. (1993), fluorescent yield are from

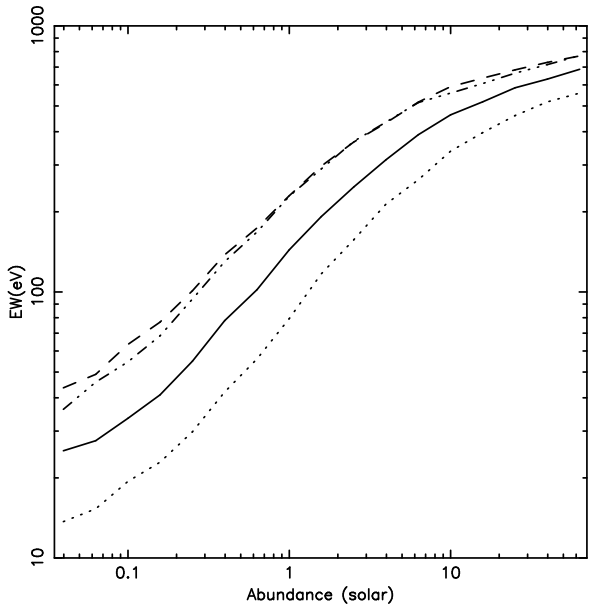

**Figure 3.** Equivalent width of the fluorescent iron K$\alpha$ line as a function of the iron abundance. The solid line shows the case in which all light elements are present with their solar abundances (as defined by Morrison & McCammon 1983). The dotted and dashed lines show the cases where the light elemental abundances are doubled and halved respectively. The dot-dashed line shows the corresponding result when all abundances are referred to the cosmic values of Anders & Grevesse (1989), with the light metals present at those cosmic abundances.

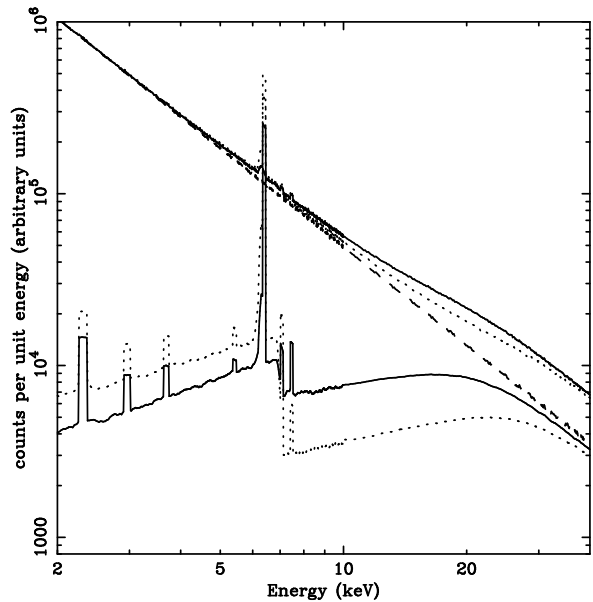

**Figure 4.** Reflected spectrum from a cold slab of material viewed at an inclination of 20 degrees as computed using Monte Carlo methods. The dashed line shows the incident power law continuum with photon index $\Gamma = 1.9$. The lower solid line is the reflected spectrum expected from a cold slab of material with solar abundances as defined by Morrison & McCammon and which subtends $2\pi$ at the X-ray source. The upper solid line is the observed spectrum in such a case (i.e. direct emission plus reflected emission). To demonstrate abundance effects on the reflected continuum, the dotted lines represent the corresponding quantities when the abundances are taken to be those of Anders & Grevesse with a three-fold enhancement in the abundance of iron (giving an iron fluorescent line with an equivalent width of $\sim$ 400 eV). It can be seen in the latter case how the reflection continuum in the *ASCA* band is suppressed.

Bambynek et al. (1972) and line energies are from Zombeck (1990). A semi-infinite slab geometry is assumed.

The Monte Carlo simulation allows us to compute the equivalent width of the iron K$\alpha$ line as a function of iron abundance. The results of such a computation are shown in Fig. 3. We also investigate the effect of varying other elemental abundances away from their solar values. Reducing the abundances of the light metals (i.e. lighter than iron) reduces the absorption of photons produced by iron fluorescence thereby increasing the equivalent width of the fluorescence line (for a given iron abundance). The effect of doubling/halving light metal abundances is shown in Fig. 3.

The present observations of Mrk 1040 give an equivalent width of 550±250 eV. Similar values have been found by *ASCA* in some other Seyfert 1 galaxies. Using solar abundances as defined by Morrison & McCammon (1983) and postulating that all light metals are present with these abundances, the observed equivalent width requires iron to be over-abundant by more than a factor of three. This can be reduced by postulating an under-abundance of the light metals. For example, halving the light metal abundance reduces the required iron enhancement to a factor of 1.5. Alternatively, using the more recent cosmic abundances of Anders & Grevesse (1989) for all elements, the strength of the iron line can be explained with only a small enhancement (factor of 1.5) in the iron abundance. The corresponding curve is also shown in Fig. 3. The values quoted here are lower limits on the required iron abundance: the upper limit on the equivalent width (800 eV) does not effectively constrain the maximum allowed iron abundance due to saturation effects (see Fig. 3).

As shown in George & Fabian (1991), increasing the iron abundance causes more iron K-shell absorption of the reflection continuum, which is therefore weakened considerably as the abundance rises. This point is illustrated in Fig. 4. Thus we can reconcile the presence of a strong iron fluorescence line with a weak associated reflection continuum.

The above enhancements in iron abundance are consistent with the iron over-abundance required by model E above for the soft spectral complexity. Thus, a self-consistent picture can be constructed (at least as far as the present X-ray data is concerned) in which the iron fluorescence line originates from X-ray reflection by rapidly moving cold material (such as an accretion disk) and the soft spectral complexity is the result of line-of-sight absorption by material with the same (non-solar) abundances. However, models for the soft complexity discussed in Section 3.3.1, together with X-ray reflection from an accretion disk with non-solar abundances provide an equally acceptable description of the present data.

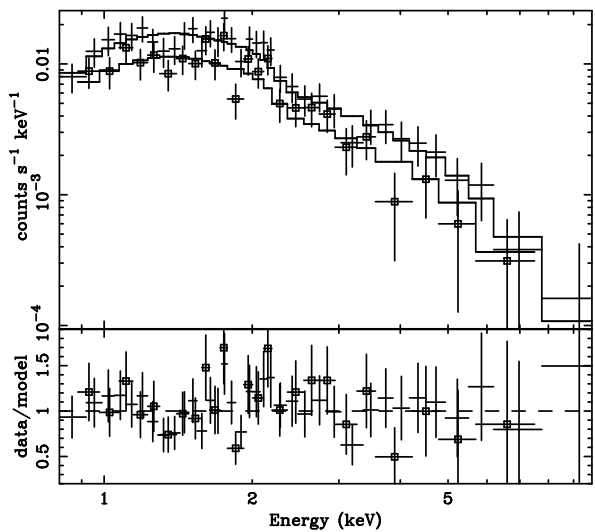

**Figure 5.** *ASCA* GIS2 (plain crosses) and GIS3 (crosses with squares) spectra of MS 0225.5+3121. $1-\sigma$ errors are shown. Also shown is a model consisting of a simple power law of photon index $\Gamma = 2.03$ absorbed by the cold Galactic column density of $N_{\rm H} = 6.9 \times 10^{20}$ cm$^{-2}$. The bottom panel shows the deviations of the data from the model.

## 4 TEMPORAL AND SPECTRAL ANALYSIS OF MS 0225.5+3121

### 4.1 Temporal analysis

Only GIS data were taken for MS 0225.5+3121 due to the restricted field of view of the SIS. The resulting GIS lightcurves show no evidence for variability during the 12 hour period of the observation. However, the variability would have to be extreme to be detected given the small count rate for this source. *Einstein* found a 0.3–3.5 keV luminosity of $2.1 \times 10^{43}$ erg s$^{-1}$ (Maccacaro et al. 1991). Extrapolating the spectral models presented below to 0.3 keV, the inferred 0.3–3.5 keV luminosity during the *ASCA* observation is comparable to the *Einstein* value (although accurate comparison is not possible due to the uncertainties in extrapolating the GIS spectra down to 0.3 keV). Thus, the data are consistent with the source being constant between the two observations as well as during each observation.

### 4.2 Spectral analysis

Spectral fitting was performed using both GIS spectra simultaneously. A simple power law fit modified by Galactic absorption ($N_{\rm H} = 6.9 \times 10^{20}$ cm$^{-2}$; Stark et al. 1992) provides a good description of the data ($\chi^2 = 52$ for 60 dof). The best-fitting photon index is $\Gamma = 2.03^{+0.11}_{-0.12}$, compatible with the canonical photon index of 1.9 from Nandra & Pounds (1994). The fit of this model to the data is shown in Fig. 5. Allowing the cold absorbing column density to be a free parameter does not improve the fit to the data ($\Delta\chi^2 < 1$) and allows an upper limit of $N_{\rm H} = 1.9 \times 10^{21}$ cm$^{-2}$ to be placed on any intrinsic absorption [Note that this is a weak upper limit due to poor low-energy response of the GIS.] There are insufficient counts to usefully constrain any possible iron emission line at 6.4 keV: an upper limit (at 90 per cent confidence) of $\sim 1$ keV to the equivalent width of any such line can be set.

Examination of Fig. 5 reveals the presence of a possible spectral feature at $\sim 1.3$ keV (similar features at $\sim 1.9$ keV and $\sim 3.1$ keV are below the spectral resolution of the GIS and thus must be considered statistical). The addition of an absorption edge gives a threshold energy of $1.26 \pm 0.14$ keV and an upper limit (at the 90 per cent confidence level) to the optical depth at threshold of 0.74. However, the F-test reveals that the addition of such an edge is not statistically significant (giving only $\Delta\chi^2 = 3$ for two additional degrees of freedom). Data containing more counts are required to examine the reality of this feature.

## 5 SUMMARY

Temporal and spectral X-ray studies of Seyfert galaxies provide valuable clues as to the geometry and physical state of the material in the central regions of AGN. The slope and variability of the underlying power law has direct bearing on the primary high energy emission mechanisms whereas deviations from the power law (at least over the *ASCA* band of 0.4–10 keV) can be interpreted as being due to X-ray reprocessing by matter in the immediate vicinity of the central engine. In an attempt to study such reprocessing, we present *ASCA* observations of the Seyfert 1 galaxies Mrk 1040 and MS 0225.5+3121. Neither object displayed variability during the observation, although the flux from Mrk 1040 has decreased by a factor of 4 since the *EXOSAT* observation 10 years ago.

Mrk 1040 shows spectral complexity both below 0.8 keV and between 6–7 keV. The latter is readily interpreted as fluorescent K$\alpha$ line emission from iron in a low ionization state (at an energy 6.4 keV in the rest frame). Such line emission is expected when a strong X-ray flux illuminates cold material, the so-called X-ray reflection model. The data show a broad line (with FWHM of 16 000–70 000 km s$^{-1}$) thereby suggesting the emission originates close to the compact object (maybe in the form of an accretion disk). The large equivalent width of the observed line ($550 \pm 250$ eV) requires us to postulate non-solar abundances. However, the equivalent width of this line alone does not allow us to uniquely specify the iron abundance since the abundance of the principle absorbers (e.g. oxygen) can also have a large effect on the observed iron line. The soft spectral complexity requires either a strong soft excess together with a large intrinsic column of neutral absorbing gas, or a more complex absorber (i.e. neutral gas which only partially covers the primary X-ray source, neutral gas with highly non-solar abundances or an ionized absorber). These possibilities are discussed in Section 3.3.

Due to the restricted SIS field of view, only GIS data are available for MS 0225.5+3121. Consequently, an investigation of any low energy (1 keV) spectral complexity is not possible. The GIS data are well described by a simple power law model with photon index $\Gamma = 2.03^{+0.11}_{-0.12}$.

**ACKNOWLEDGMENTS**

We thank Niel Brandt, Roderick Johnstone, Paul Nandra and Dave White for valuable discussions throughout the course of this work. We are also extremely grateful to Lorella Angelini, Ken Ebisawa and Don Jennings of the Goddard Space Flight Center (GSFC) for their help with the FTOOLS software package. This research has made use of the NASA/IPAC Extragalactic Database (NED) which is operated by the Jet Propulsion Laboratory, Caltech, under contract with the National Aeronautical and Space Administration.